# On plastic deformation and fracture in Si films during electrochemical lithiation/delithiation cycling

Siva P V Nadimpalli[1], Vijay A. Sethuraman[1], Giovanna Bucci[1], Venkat Srinivasan[2], Allan F. Bower[1], Pradeep R Guduru**,[1]

[1]School of Engineering, Brown University, Providence, Rhode Island 02912, USA

[2]Environmental Energy Technologies Division, Lawrence Berkeley National Laboratory, Berkeley, California 94720, USA

**Corresponding author, email: pradeep_guduru@brown.edu, Tel: +1 401 863 3362, Fax: +1 401 863 9009

**Abstract**

An *in situ* study of deformation, fracture, and fatigue behavior of silicon as a lithium-ion battery electrode material is presented. Thin films (100-200 nm) of silicon are cycled in a half-cell configuration with lithium metal foil as counter/reference electrode, with 1M lithium hexafluorophosphate in ethylene carbonate, diethylene carbonate, dimethyl carbonate solution (1:1:1, wt.%) as electrolyte. Stress evolution in the Si thin-film electrodes during electrochemical lithiation and delithiation is measured by monitoring the substrate curvature using the multi-beam optical sensing method. The stress measurements have been corrected for contributions from residual stress arising from sputter-deposition. An indirect method for estimating the potential errors due to formation of the solid-electrolyte-interphase layer and surface charge on the stress measurements was presented. The films undergo extensive inelastic deformation during lithiation and delithiation. The peak compressive stress during lithiation was 1.48 GPa. The stress data along with the electron microscopy observations are used to estimate an upper bound fracture resistance of lithiated Si, which is approximately 9-11 J/m$^2$. Fracture initiation and crack density evolution as a function of cycle number is also reported.

## 1. Introduction

Silicon is recognized to be a promising alternative as negative electrode for the next generation of high energy density Li-ion batteries due to its high charge capacity [1] of 3579 mAh/g compared to that of graphite (372 mAh/g). However, large volume changes of Si during insertion and extraction of lithium are known to cause fracture and capacity loss, limiting the widespread use of Si electrodes in Li-ion battery applications [2]. For instance, electrodes made of bulk Si film and micron sized Si particles suffer severe capacity fade in very few lithiation/delithiation cycles [3,4]. Electrode fracture not only leads to mechanical degradation but also contributes to chemical degradation as electrolyte reduction occurs on fresh fracture surfaces as solid electrolyte interphase (SEI) layer, which leads to additional capacity loss [5]. Several strategies have been proposed to alleviate fracture and to improve cyclic performance of Si electrodes. For example, Chan *et al.* [6] used electrodes made of Si nano-wires and obtained stable capacities of approximately 3200 mAh/g for almost 20 cycles. Graetz *et al.* [7] obtained stable capacities of 2000 mAh/g for 50 cycles by using ~100 nm thick amorphous Si (a-Si) film electrode. Although these electrode designs improved the cyclic performance, they do not completely eliminate the fracture problem. Obrovac et al. [8] showed that when crystalline Si reacts with lithium it becomes an amorphous $Li_xSi$ phase, and if lithiation continues below 0.05 V vs. $Li/Li^+$, the amorphous phase crystallizes into $Li_{15}Si_4$ phase resulting in a maximum theoretical capacity of 3579 mAh/g. Such phase transformations generally induce significant stresses in the material and speed up the damage process [8].

It is evident that electrode deformation and the associated stresses play a critical role in damage initiation and evolution. There have been several models of stress evolution, diffusion, and fracture of electrodes in recent literature [9-13]. There have also been a limited number of experimental investigations in recent literature on characterizing mechanical properties of lithiated silicon [14-20]. Lee et al. [14] measured the stress response in a 35 nm amorphous Si (a-Si) film and noted that the stresses were compressive during insertion of Li and tensile during extraction of Li from Si. However, it was primarily a qualitative study and the actual magnitude of stresses was not reported. Sethuraman et al. [15] performed in situ stress measurements using the substrate-curvature technique in 250 nm a-Si films during lithiation and delithiation and reported that the a-Si undergoes elastic-plastic deformation and experiences a peak stress of 1.75 GPa. In a recent study, Zhao et al. [12] studied deformation of lithiated Si using first principle calculations based on density functional theory and postulated that the initial elastic deformation in lithiated Si was accommodated by insertion of Li atoms into interstitial sites in the a-Si network without rupturing any Si-Si bonds. Beyond the elastic regime, the Li insertion was accommodated by rearrangement of Si-Si bonds with Li-Si bonds which allowed the $Li_xSi$ electrode to undergo plastic deformation. Sethuraman et al. [15] also showed that the energy dissipation associated with plastic deformation of lithiated Si constitutes a significant portion of the total energy loss in subjecting the half-cell to a full lithiation/delithiation cycle. In subsequent studies, they measured the biaxial modulus of $Li_xSi$ [16] which decreased significantly with state of charge (i.e., Li concentration) and also demonstrated that the stress will have a significant effect on the electrode potential [17]. The latter phenomenon was also discussed recently by Sheldon et al. [18].

It is well known that the electrochemical processes generally induce surface stresses due to the presence of excess charge on electrode surfaces [21,22], and understanding the contribution of the surface stress to the substrate curvature is necessary in estimating the errors associated with stress measurements using the in situ substrate curvature method. Moreover, during lithiation, at potentials below ~0.8V vs. $Li/Li^+$, the electrolyte is reduced on the Si surface, resulting in formation of the SEI layer which grows in thickness, preventing further electrolyte reduction. Investigating the influence of the SEI layer formation on curvature change would be helpful in understanding the contribution to the electrode stress from lithium intercalation versus SEI formation.

Si electrodes, regardless of the configuration (i.e., Si nano-particles, nano-films, and nano-wires), are subjected to repeated stress cycles during electrochemical charge/discharge cycles. Such stress cycling eventually leads to crack initiation, propagation, and electrode failure. Several studies on the cycling behavior of Si electrodes exist in literature. For example, Maranchi et al. [23] have studied the cycling behavior of a 250 nm a-Si film on Cu and found out that the film developed cracks and separated into islands. Kasavajjula et al. [1] provide a review of several studies on the cyclic behavior of Si and Si-based electrodes. However, the focus of all these studies was mainly to understand the electrochemical performance of the electrodes during cycling process, and not on their mechanical response. To be able to design durable Si electrodes that can sustain cyclic loading, it is essential to understand the stress evolution during electrochemical cycling and damage mechanisms associated with them. Though the electrode fracture and the associated capacity loss are widely acknowledged as common problem, there appear to be no studies on quantitative characterization of the fracture behavior of the lithiated Si.

Building on the preliminary experiments reported by Sethuraman *et al.* [15], the primary objectives of this work are (i) to carry out a more refined set of experiments on the elastic-plastic behavior of amorphous Si (*a*-Si) films upon electrochemical lithiation and delithiation (ii) to study evolution of electrode stresses and cracking as a function of charge/discharge cycles, and (iii) to estimate an upper bound fracture resistance of lithiated Si. To this end, Si thin-film electrodes were prepared by depositing *a*-Si films of 100 nm and 200 nm thickness on elastic substrates and cycled under a range of electrochemical conditions. The residual stresses in the *a*-Si film due to sputter-deposition process were measured before lithiating the samples. Stress evolution in *a*-Si films during lithiation and delithiation was measured by monitoring the substrate curvature using a multi-beam optical sensing (MOS) system [15]. Curvature changes due to surface stress and SEI formation on Cu films were measured by cycling Cu against Li foil. These measurements were then used to deduce estimates of curvature change due to SEI formation on Si electrodes. The stress data along with the scanning-electron microscopy (SEM) images and film properties were used to calculate an upper bound on the fracture resistance of lithiated Si. A systematic investigation of damage evolution was carried out through *ex situ* SEM analysis; preliminary data on the low cycle fatigue failure of the $Li_xSi$ thin films is also presented.

## 2 Experimental Procedures

### 2.1 Electrode Fabrication and Electrochemical Cell

Electrodes were prepared by depositing 20 nm of Ti, 300nm of Cu, and Si films on (111) single crystal Si substrates of 400 μm thickness and 50.8 mm diameter, as shown in Fig. 1 inset. Prior to any film deposition, approximately 300 nm thick $SiO_2$ layer was thermally grown on all sides of the Si substrates in order to create a barrier for Li diffusion and to isolate the Si substrate from all electrochemical reactions (which was verified experimentally ). It should be noted that the role of the Si substrate is to serve as an elastic substrate which undergoes curvature change in response to stress in the Si film deposited on it; as such, it can be replaced by any other elastic material that does not participate in the electrochemical reactions. The Ti layer acts as an adhesion layer between the $SiO_2$ and the Cu current collector film. The Ti and Cu films were deposited by e-beam evaporation whereas the Si films were deposited by RF-magnetron sputtering at 180 W power and less than 2 mTorr Ar pressure (Lesker PVD Lab-18 from Kurt J. Lesker Inc., PA, USA). The sputtered Si films under these conditions are known to be amorphous, which was confirmed by X-ray diffraction analysis of the deposited films [5].

To study stress and damage response under cyclic loading, 100 nm thick Si films were deposited on double side polished (DSP) substrates; for fracture studies, 200 nm thick films were deposited on the rough side of single side polished (SSP) substrates; SSP substrates have improved adhesion with the films and minimize the possibility of delamination during electrochemical lithiation/delithiation cycles. To investigate the influence of SEI layer formation and any possible reaction between Li and Cu layer on the mechanical and electrochemical measurements, samples with only Cu films were also prepared. The electrochemical cell used in the experiments is illustrated schematically in Fig.1, which was assembled in an argon-filled glove box (maintained at 25 °C and less than 0.1 ppm of $O_2$ and $H_2O$). The *a*-Si film was used as the working electrode and a lithium metal foil was used as the reference/counter electrode. The electrodes were separated by a Celgard polymer separator and submerged in an electrolyte solution of 1 M $LiPF_6$ in 1:1:1 ratio of ethylene carbonate (EC): diethyl carbonate (DEC): dimethyl carbonate (DMC). A glass window enables substrate curvature measurement through the MOS system as shown in Fig. 1. All experiments were conducted inside an argon-filled glove compartment.

## 2.2 *In situ* Measurement of Film Stress

Stress evolution in the *a*-Si films during lithiation and delithiation was measured by monitoring the substrate curvature with the MOS setup (k-Space Associates, Dexter, MI) illustrated in Fig.1. The system consists of a laser source that generates a single focused beam, two etalons arranged orthogonally to each other to generate a 2 X 2 array of beams, and a CCD camera to capture the reflected beam-array from the sample surface. The relative change in the distance between the laser spots on the CCD screen gives the sample curvature, $\kappa$, as

$$\kappa = \frac{\cos\phi}{2L}\left\{\frac{D_o - D}{D_o}\right\}, \quad\quad\quad\quad\quad\quad \text{Eq. 1}$$

where $D$ is the distance between the laser spots and $D_o$ is its initial value. $\phi$ is the reflection angle of the beam, and $L$ is the optical path length from sample to the CCD camera. The factor $cos\phi/2L$, known as mirror constant, is specific to a setup and is obtained by calibrating the system in Fig.1 with a reference mirror of known curvature in the sample plane. The 2 X 2 array of reflected spots enables curvature measurement in two orthogonal directions. The biaxial film stress is related to substrate curvature through the Stoney equation [24,25],

$$\sigma = \sigma_r + \frac{E_s t_s^2 \kappa}{6 t_f (1-v_s)} \quad\quad\quad\quad \text{Eq.2}$$

where $E_s$, $v_s$, and $t_s$ are Young's modulus, Poisson ratio, and thickness of the substrate, respectively. $\sigma_r$ is the residual stress in the *a*-Si film due to sputter deposition; $t_f$ is the film thickness which evolves continuously during lithiation and delithiation processes. Beaulieu *et al.* [26] showed by *in situ* measurements that the thickness (and volume) of a-Si increases linearly with the state of charge (SOC) as,

$$t_f = t_f^o(1 + 2.7z) \quad\quad\quad\quad\quad\quad \text{Eq.3}$$

where $t_f^o$ is the initial film thickness, $z$ is the state of charge (SOC) which changes from 0 to 1. $z=1$ corresponds to a capacity of 3579 mAh/g and a volumetric strain of 2.7.

From Eq.2, note that the film stress is independent of film properties; hence, the curvature change represents the state of stress in the film even when the film undergoes plastic deformation, phase changes, or any other nonlinear phenomenon [25]. However, the equation can lead to errors if the film develops cracks with spacing of less than about 50 times to the film thickness. For cracked films, a corrected form of the Stoney can be used, in which the curvature change due to film cracking is given by [25],

$$\Delta \kappa \approx -\kappa_{nc} \frac{l}{p} \tanh\left(\frac{p}{l}\right) \quad \text{............................ Eq. 4}$$

where $\kappa_{nc}$ is the curvature of the film with no cracks and $p$ is the average spacing between cracks. $l$ is a length scale that depends on the elastic mismatch between the film and the substrate; for the film-substrate system used in this investigation (with the properties reported in Table 2), it is approximately $3t_f$. In general, the errors associated with film cracking can be important mainly when the film is subjected to tensile stress, because cracks in an ideal elastic material will be closed under a compressive stress, making the cracked film act like a continuous film.

## 2.3 Electrochemical Measurements

The electrochemical cells were cycled using a Solartron 1470 E potentiostat; the 200 nm samples were lithiated galvanostatically at 5 µA/cm$^2$ with a lower cut-off potential of 0.05 V *vs.* Li/Li$^+$, delithiated at the same current density until a potential of 1.2 V *vs.* Li/Li$^+$. Delithation was continued potentiostatically at 1.2 V *vs.* Li/Li$^+$ until the current becomes negligible (*i.e.*, < 0.05 µA/cm$^2$). The last step ensures nearly complete delithiation of the films. The cells were kept at open circuit for 5 minutes between consecutive steps. After one lithiation/delithiation cycle, these cells were dissembled and SEM analysis was conducted on the electrodes. To study the cycling behavior of *a*-Si film, 100 nm thick *a*-Si films were cycled 48 times between upper and lower potential limits of 0.6 V *vs.* Li/Li$^+$ and 0.05 V *vs.* Li/Li$^+$ respectively at a constant current density of 8.75 µA/cm$^2$. The delithiation potential was limited to 0.6 V *vs.* Li/Li$^+$ to delay film cracking. To investigate the crack evolution behavior, some cells were dissembled after 6, 15, and 48 cycles and SEM analyses were carried out.

Measurement of stress vs. capacity is influenced by the formation of SEI layer in at least two ways. First, SEI layer could itself be under stress, as reported recently by Mukhopadhyay et al. [27] in the context of graphite lithiation. Further, since there is no easy way to separately measure the fraction of Li flux being inserted into Si and that consumed in forming the SEI layer, particularly during the first cycle, it is not possible to measure the true the state of charge (SOC) of the film. These issues are addressed by carrying a set of experiments in which thin Cu films (50 – 100 nm thickness) are held at constant potential against Li metal, allowing a mature SEI layer to grow on them while measuring the substrate curvature. The potential hold is maintained until the current density becomes negligible (< 0.05 µA/cm$^2$). It should be noted that although the nature of SEI that forms mainly depends on the type of electrolyte, the substrate can also influence the SEI that forms on it. However, measurement of Li consumption per unit area and the substrate curvature due to SEI formation allow us to estimate the uncertainty in stress and SOC measurements due to SEI formation.

## 3 Results and Discussion

## 3.1 Residual stress measurement in Si films

In the experiments reported by Sethuraman *et al.* [15], no attempt was made to measure the residual stress in the *a*-Si films. Since the curvature change in Stoney equation (Eq. 2) reflects only the change in film stress but not the actual stress magnitude, ignoring the residual stress could introduce errors in the stress values reported in those measurements. In general, thin film deposition is performed at elevated temperatures, and when the film/substrate system is cooled to room temperature, depending on the thermal expansion properties of the film and substrate, residual stresses develop in the film. Residual stresses could also develop due to coalescence of individual islands during thin film growth [28]. Residual stress in the film was measured in all experiments reported here by measuring the substrate curvature (with the setup shown in Fig.1) immediately before and after *a*-Si deposition and using the Stoney Equation [Eq. 2 with only the second term on the right hand side]. A specially designed sample holder was used to ensure that the curvature measurements were made at the same spot and sample orientation of the substrate in order to minimize errors due to any non-planarity of the substrate. The residual stress values measured in four representative samples are shown in Table 1. It can be noted from the table that the residual stress in *a*-Si films deposited by sputtering process can be as high as 0.5 GPa and therefore, cannot be neglected.

### 3.2 Influence of Surface Stress and SEI layer formation on Substrate Curvature Measurements

It is well-known in electrochemistry literature that surface charge density can induce surface stress (also known as electrocapillary effect) [21,22], which can influence the curvature measurements. Moreover, it is possible that the formation of the SEI layer can influence the substrate curvature change during electrochemical cycling, which can happen in two ways: (i) SEI layer could itself be under stress as reported by Mukhopadhyay et al. [27] for graphite electrodes and (ii) SEI layer could influence the charge density and modify the electrocapillarity-induced surface stress. In order to investigate the contribution of these factors on the curvature measurements and hence the film stress measurements, Cu films on elastic substrates are electrochemically cycled, while measuring the substrate curvature in a setup shown in Fig. 1.

Fig. 2(a) shows potential history and curvature evolution when a 50 nm Cu film is "lithiated" galvanostatically (current density 5 µA/cm$^2$) until 0.05 V vs. Li/Li$^+$ and held at that potential subsequently. The curvature evolves quickly during the initial 10 hrs, corresponding to a stress-thickness of about 22 Pa m and subsequently evolves slowly, reaching a value of about 24 Pa m. Note that the stress evolution is not steady, the fluctuations indicate that the process of SEI formation is dynamic, with the composition of the SEI layer and its mechanical properties continuing to evolve with time. The stress appears to be relatively steady beyond 48 hours. A thorough study of SEI layer formation mechanisms, chemical and physical structure, film morphology and its mechanical properties is required to fully understand how it contributes to curvature change shown in Fig. 2(a), which is beyond the scope of this work. However, measurements in Fig. 2(a) allow us to estimate the uncertainty in stress measurements in the electrode films based on substrate curvature method as follows. Note that the stress-thickness evolution was monitored in Fig. 2a over a duration of 60 hours; a typical lithiation experiment of a 100 nm Si film in the present investigation was done at C/14 rate, at the end of which the film thickness is *ca.* 264 nm (charge capacity of ~2180 mAh/g). Taking the stress-thickness value due to SEI at 14 hr to be 21 Pa m (Fig. 2a), it would result in overestimation of stress in the Si film by about 79 MPa, which represents an uncertainty of 8.4% in the measured stress of approximately 1 GPa in the film at that SOC. Although this is a relatively small error for a 100 nm (or thicker) film, it could represent a 44% error in a 20 nm film. It is worth pointing out that the error estimates based on the stress-thickness measurements due to SEI formation on Cu films are assumed to be valid for Si films as well, which is a reasonable assumption since it is generally thought that the electrode material acts as a catalyst for electrolyte reduction reactions, whereas the SEI products are primarily a function of the electrolyte composition. Recent studies by Nie et al. [29, 30] show that the SEI products formed on Si and Graphite electrodes are similar; also, the products on these electrodes are not too different from the ones that formed on Li metal electrodes reported by Aurbach et al. [31]. Hence, the reader should note that although the data based on the SEI formed on Cu can give good error estimates in stress measurements due to SEI formation on Si, the limitation of the study is that they are indirect estimates and cannot be considered as an absolute correction.

It would be instructive to consider the origins of stress in the SEI layer, although a full investigation is beyond the scope of the present investigation. It has been suggested that Li intercalation in the SEI layer could be a source of stress [27]. Curvature change during lithiation could have partial contribution from the electrode potential through the electrocapillary effect. To investigate this effect, electrodes were held at 0.6 V *vs.* Li/Li$^+$ until the SEI current becomes negligible and were then subjected to step changes in potential as shown in Fig. 2b. The step potential changes result in instantaneous changes in curvature, providing evidence of how potential affects surface stress and thus the curvature. The corresponding curvature change, however, is relatively small, *ca.* 3 Pa m, which is consistent with the range of surface stress values that are generally reported for typical electrochemical processes [22]. Thus, excluding the surface stress effects, the stress-thickness due to the SEI layer alone is around 20 Pa m. In order to convert it to a "stress" in the SEI layer, one would need to know the thickness of the SEI layer. The SEI layer formation mechanisms, its physical structure, and its mechanical properties are not well understood and they constitute an active area of current research in the field. There is a large variation in the reported values of thickness and properties of the SEI layer. For example, Wu *et al.* [32] observed that approximately 110 nm thick SEI layer was formed on Si nano-tubes after electrochemical cycling, whereas Bhattacharya *et al.* [33] showed that approximately 1 µm thick SEI layer formed on graphite electrodes during a voltammetry experiment. If 110 nm is assumed for SEI layer thickness in Fig. 2a, the stress in the SEI layer would be approximately 218 MPa, which is very high for an SEI layer that consists of reduction products of organic solvents. At the same time, studies [33,34] also showed that the SEI layer could contain layers of stiff carbonate, which could readily sustain large stresses. Although a separate investigation is required to understand how the SEI layer can sustain such high stresses, it is worth noting that stress measurements in the SEI layer could be a useful diagnostic tool to complement the existing techniques used to investigate the structure and composition of SEI.

### 3.3 Elastic-Plastic Deformation of Lithiated Silicon

Figs. 3 (a) and (b) show the potential and stress response of 100 nm *a*-Si film (Sample 4 of Table 1) during lithiation and delithiation as a function of capacity, where the charge contribution to the SEI layer formation was minimized during the stress measurements. The experiments were carried out in two steps: In the first step, represented by the dashed line in Fig.3 (a), the sample was lithiated at a constant current density until the potential reached 0.6 V *vs.* Li/Li$^+$, where it was held constant until the current became negligible (~0.1 µA/cm$^2$). The constant potential value is chosen to be below the reduction potential of the electrolyte (~ 0.8V *vs.* Li/Li$^+$) to allow a partial SEI layer to grow; at the same time, high enough to minimize lithiation of the Si film. The latter was verified by subsequently holding the electrode potential at 1.2 V *vs.* Li/Li$^+$ and measuring the Li recovered from the electrode, which was less than 2 mAh/g. The samples, with the SEI layer partially grown on them, are then lithiated and delithiated at a constant current density of 5 µA/cm$^2$, while monitoring the substrate curvature change, which are represented as solid lines in Fig. 3(a). Since a partial SEI layer is formed *a priori*, the capacity measurement represents the Li concentration in Si more accurately compared to those reported by Sethuraman et al. [15]. For the 100 nm film used in these experiments, the Li lost to SEI formation is equivalent to an apparent capacity of about 800 mAh/g, which by itself is a significant correction, although SEI layer will continue to form below 0.6 V. However, it should be emphasized that SEI will continue to form at potentials below 0.6 V, which will contribute uncertainty in SOC measurement.

The solid curve in Fig. 3a shows that the cell potential drops sharply to below 0.5 V *vs.* Li/Li$^+$ at the beginning and decreases slowly to 0.05 V *vs.* Li/Li$^+$ afterwards, indicating that Si lithiation, *i.e.*, insertion of Li into *a*-Si film, begins approximately below 0.5 V *vs.* Li/Li$^+$. Absence of any constant potential regions in the solid curve suggests that there were no two-phase regions in the film during alloying (lithiation/delithiation) process and the addition of Li to amorphous Si results in amorphous Li$_x$Si alloy (a homogeneous solid solution), consistent with Ref.[8]. During lithiation, the substrate constrains the film from in-plane (*x-y* plane) expansion, resulting in equi-biaxial compressive stress in the film, Fig. 3(b); the film can expand in the out-of-plane (*z*) direction only. Note that the stress values are corrected for residual stress in the film (the stress is non-zero at zero capacity in Fig. 3b). Initially, the compressive stress increases linearly with lithiation, which corresponds to the elastic response of the film, and becomes non-linear at approximately 140 mAh/g, indicating the onset of plastic deformation of lithiated Si. The stress reaches a peak value of -1.48 GPa at 150 mAh/g and decreases to -1 GPa at 700 mAh/g. Subsequent lithiation appears to result in only small changes in stress, gradually decreasing to about -0.94 GPa at a capacity of 2180 mAh/g. Using first principle calculations based on density functional theory, Zhao *et al.* [12] showed that the *a*-Si film could accommodate plastic deformation without further increase in the stress level, as in Fig. 3(b), due to the rearrangement of Si-Si bonds with Li-Si bonds. Since the film continues to be in a state of plastic deformation, the measured stress history in Fig. 3(b) can be viewed as the evolving yield stress of lithiated Si as a function of Li concentration. During delithiation (see Fig. 3(b)) the substrate prevents the contraction of the film which induces a tensile stress. As a result, the stress changes rapidly and becomes tensile with a small change in capacity of ~100 mAh/g, which represents elastic unloading-reloading of the film. Upon subsequent delithiation, the film begins to deform plastically in tension at a tensile stress of about 0.4 GPa, which is lower than the yield stress under compression for the same capacity. The film continues to deform plastically at a tensile stress of 0.4 GPa and the stress begins to increase at around a capacity of 1100 mAh/g, which mirrors a similar decrease during lithiation between 150 and 700 mAh/g. The schematic in Fig.3c shows graphically the evolution of curvature, Li concentration, and film thickness at various stages of lithiation/delithiation indicated in Fig.3b. These observations are qualitatively consistent with those of Sethuraman *et al.* [15], although there are differences in capacity and stress magnitudes. For example, the peak compressive stress measured here is smaller than the previously reported values of 1.75 GPa by Sethuraman et al. [15] and 2 GPa by Zhao et al. [12]. We believe that the results reported in this work are more accurate than those reported in the earlier work of Sethuraman *et al.* because of residual stress measurement and better uncertainty estimation due to SEI layer [15].

## 3.4 Upper Bound Fracture Energy of Lithiated Silicon

Figs. 4(a) and 4(b) show the potential and stress evolution, respectively, in a 200 nm thick *a*-Si film during lithiation and delithiation as a function of capacity, without accounting for SEI formation. The films were deposited on the rough side of SSP wafers to improve adhesion. Note that the rough side of SSP wafers have a "terrace" texture, consisting of flat terraces of up to a few tens of microns in dimension, separated from each other by a small height difference (up to a micron) in the *z*-direction. As expected, the onset of plastic deformation in this case occurs at a higher capacity. However, the peak compressive stress in this case, 1.35 GPa (Fig. 4(b)) is very close to that in Fig. 3(b); the small difference could be accounted for by the terrace-texture of the film. An SEM image of the electrode surface (of Fig. 4(a)) after the first complete lithiation- delithiation cycle is shown in Fig.4c, which reveals that the Si film developed a network of channel cracks. Although the exact value of stress at which the cracks initiated could not be directly observed, the peak tensile stress of 1.21 GPa at which the curvature drops represents an upper bound. When the energy release rate, *G*, exceeds the fracture resistance of the film, *Γ*, cracks spread by channeling as seen in Fig. 4c. The surface roughness of the substrate, especially the edges of the terraces (Fig. 4c) possibly provide crack nucleation sites or flaws. By assuming the onset of cracking at the peak tensile stress, *i.e.*, 1.21 GPa (Fig.3b), an upper-bound, the fracture energy per unit area, *Γ*, of lithiated Si can be calculated.

Nakamura and Kamath [35] analyzed the problem of channel cracking in a film on an elastic substrate using a three-dimensional finite element method and showed that as soon as the flaw size reaches a value close to film thickness, $t_f$, the energy release rate becomes independent of crack length. Energy release rate for such a steady state channeling, as in Fig.4c, can be calculated by [36,37],

$$G = \frac{\pi}{2} \frac{(1-\nu_f^2) t_f \sigma^2}{E_f} g(\alpha, \beta) , \quad \text{................................ Eq. 5}$$

where $\nu_f$ and $E_f$ are Poisson ratio and Young's modulus of the film, respectively. $\sigma$ is the critical stress where channel cracks propagate (assumed to be the peak stress in Fig. 4b), and the function *g* depends on the elastic mismatch between the film and substrate through Dundur's parameters *α* and *β*. According to [37], the dependence of function *g* on *β* can be neglected; for the film/substrate properties of Table 2, α = -0.36 and the function *g* attains a value of 1.03. By substituting in Eq. 5, $t_f$ = 260 nm, calculated according to Eq.3, at the onset of cracking (*i.e.*, at 390 mAh/g); $E_f$ value (*i.e.*, Young's modulus of $Li_{0.4}Si$ alloy from [16,20, 39]) of Table 3; and $\sigma_c$ = 1.21 GPa, the fracture energy of lithiated Si was calculated and presented in Table 3. It should be noted that the thickness estimate of 260 nm was not corrected for SEI formation; however, the error due to this will be negligible for low SOC (such as at 390 mAh/g). The fracture energy and the mode-I fracture toughness, $K_{Ic}$, values of single crystal Si are also presented in the table for comparison. It is interesting to see that the upper-bound fracture resistance *Γ* = 9-11 J/m$^2$ of amorphous $Li_{0.4}Si$ is not very different from the fracture resistance of crystalline Si 5-14 J/m$^2$, in spite of the fact that the fundamental mechanisms of deformation could be very different in the two materials. The fracture energy value for lithiated Si obtained from this study, although an upper bound estimate, could be useful in modeling and/or design of durable Li-ion battery electrode architectures.

**3.5 Low Cycle Fatigue of *a*-Si Films during Electrochemical Cycling**

It is evident from Figs. 3 and 4 that *a*-Si deforms elastic-plastically under compressive stress during lithiation and tensile stress during delithiation. Consequently, when batteries containing Si-based electrodes are subjected to several electrochemical (charge/discharge) cycles, portions of the electrodes, irrespective of their geometry (*e.g.*, nano-wires, nano-particles, and nano-films), undergo compression-tension cycles. In general, materials that are subjected to such repeated large elastic-plastic stress cycles in service fail by low cycle fatigue [43], and understanding how the electrode stresses evolve during electrochemical cycling not only reveals how electrode degradation progresses, but also provides the necessary information for designing durable battery electrodes. Existing studies on cyclic behavior of Si [23] have contributed to the understanding of the electrochemical performance of the electrodes, but do not provide insights on stress and damage evolution during electrochemical cycling. In this study 100 nm thick *a*-Si films were subjected to multiple electrochemical (lithiation/delithiation) cycles and the corresponding stress and damage evolution was monitored.

Figs. 5a and Fig. 6a show the potential response (during the first six cycles) of a 100 nm *a*-Si film (sample 3 of Table 1) as a function of time and capacity, respectively, and Fig. 7a shows the data for a few selected cycles. Note from the figures that during the first cycle lithiation the cell potential drops sharply to approximately 0.4 V *vs.* Li/Li$^+$ suggesting that Si lithiation begins around 0.4 V *vs.* Li/Li$^+$; however, in subsequent cycles the Si lithiation begins at 0.25 V *vs.* Li/Li$^+$. The discrepancy is possibly due to the fact that the starting material for the first cycle lithiation is as-sputtered amorphous Si, whereas for the subsequent cycles, it is amorphous Si that has been subjected to prior lithiation. Since delithiation potential was limited to 0.6 V *vs.* Li/Li$^+$, the starting material from second cycle onwards is partially lithiated *a*-Si, which accounts for the lower lithiation potential. Fig. 7b shows that the first cycle lithiation capacity of the 100 nm Si film (Sample 3 of Table 1) was 2817 mAh/g and the corresponding delithiation capacity was 1948 mAh/g. The lithiation capacity continued to decrease with cycle number and reached a steady state value of approximately 1180 mAh/g by the 11$^{th}$ cycle; the corresponding steady state the delithiation capacity is around 1160 mAh/g. The decrease in the cell capacity could be due to continued reduction of the electrolyte and gradual accumulation of mechanical damage both of which could increase the cell impedance; the initial decrease in the capacity was mainly due to electrolyte reduction and irreversible changes in the Si electrode. These observations, from Figs. 5a, 6a, and 7a, are consistent with the previous reports on Si films, eg. [23]. Although this data is necessary to understand the lithiation mechanism and capacity fading as a function of electrochemical cycles, it is not sufficient to understand the mechanical damage and failure of electrodes.

Fig. 8 shows SEM images of as-deposited *a*-Si film and films after an increasing number of lithiation-delithiation cycles. It should be noted that the 100 nm *a*-Si film does not develop cracks for at least the first 6 cycles (*i.e.*, no mechanical damage), but develops extensive cracking and separates into islands after 48 cycles. Note that the opening of the cracks increases with cycling, and one possible explanation for this could be the interface sliding between Si and Cu which causes shrinking of islands and increased gap between them [44]. Although crack evolution could be monitored by subjecting the films to progressively increasing number of cycles and carrying out SEM imaging, it could be monitored indirectly by continuously measuring the substrate curvature since the Stoney equation is modified by the presence of film cracks.

Figs. 5b,c and 6b,c show the curvature (stress-thickness) and stress response of the 100 nm Si film (sample 3 of Table 1) as a function of time and capacity, respectively, for the first 6 cycles; Fig. 9 shows the curvature evolution for the subsequent ~40 cycles. As the stress-thickness quantity is directly proportional to substrate curvature, both these terms are used interchangeably in the following discussion. Note from Figs. 6c and 9a, that the stress and curvature response was repeatable for the first 7 cycles (9a), *i.e.*, the stress (or curvature) at a given capacity is independent of cycle number, which suggests that cracking was absent during the initial 7 cycles, consistent with the SEM observations, Fig.8b. With subsequent cycling, the area enclosed by the curvature-capacity loop decreases slightly, Fig. 9(a). For example, the curvature value at any capacity during the $8^{th}$ cycle was slightly smaller than that of the $7^{th}$ cycle. Such decrease in curvature can be interpreted as formation of a crack network, which depends on the ratio of crack spacing to the film thickness [25] given by Eq.4. Thus, a reduction in curvature in the $8^{th}$ cycle indicates the onset of cracking in the Si film cycled under the present conditions. It should be noted that the number of cycles to onset of cracking could depend on parameters such as the maximum strain in the electrode during cycling (which would be proportional to the state of charge), charge/discharge rates, and surface finish of electrodes. The mechanism of crack initiation in amorphous $Li_xSi$ is not entirely clear at present, however, an examination of the stress and microstructure evolution in Figs. 2d, 8, and 9 and some of the recent literature reports [45,46] suggest some possibilities. First, it has been observed in the present investigation and also reported in recent literature [45] that the first few lithiation-delithiation cycles result in significant surface roughening (Fig. 8e). In addition, lithiation of Si could also cause porosity according to Son *et al.* [46]. Thus, appearance of critically sized flaws in the first few cycles under extensive cyclic plastic straining is akin to low cycle fatigue in structural materials. It should also be noted that fatigue crack initiation from evolved surface features is a typical mechanism in ductile crystalline materials [43]. Micro-scale mechanisms of morphology and flaw evolution in lithiated silicon remain an open and important issue that deserves further attention.

The curvature *vs.* capacity loop (Fig. 9b) continues to shrink from $11^{th}$ to $20^{th}$ cycle indicating further electrode cracking, *i.e.*, increase in the crack density. Note from Eq.4 that the curvature will decrease if the crack spacing decreases (or crack density increases). The loop shrinking continues at a slower rate from $21^{st}$ to 28 cycles, changes minimally from $29^{th}$ to $39^{th}$ cycle, and almost ceases from $40^{th}$ to $48^{th}$ cycles. This indicates that the crack density evolved until the $29^{th}$ cycle causing fragmentation of the film into islands which become stable from the $30^{th}$ cycle onwards, resulting in stable capacities of 1160 mAh/g until $48^{th}$ cycle as in Fig.7. This suggests that the island size at the $30^{th}$ cycle was not too different from that at the $48^{th}$ cycle, which is approximately 1 μm as shown in Figs. 8c and d. Hence, a possible strategy to avoid film fracture would be to design patterned film electrodes with islands of the order of 1 μm size [44,47,48].

## 4. Summary

The main results of the work can be summarized as follows. Since stress evolution in thin electrode films during electrochemical cycling through substrate curvature method is becoming increasingly common, it is necessary to understand the possible sources of uncertainty in the measurements. In this work, we attempted to describe and correct for some of the sources of uncertainty such as residual stress, surface stress (electrocapillary effect) and contribution of SEI layer to curvature. The main conclusion is that lithiated Si begins to deform plastically at a stress of 1.48 GPa, which is lower than the value of 1.75 GPa reported by Sethuraman *et al.* [15]. Based on the curvature measurements, surface stress due to electrocapillary effect is estimated to be ~3 Pa m and that due to SEI layer can be as high as 21 Pa m. Neglecting these factors could result in improper estimates of intercalation induced stresses in Si films, especially in very thin films (< 50 nm). Based on the appearance of channel cracks in the film at the end of a lithiation-delithiation cycle and the stress measurements, an upper bound fracture energy of $Li_{0.4}Si$ was estimated to be approximately 9- 11 $J/m^2$. Continued cycling of the 100 nm thick films results in fragmented islands of about 1 μm in dimension, which could be an appropriate dimension for patterned islands for increased cycle-life.

## Ackonwledgements

The authors gratefully acknowledge funding from the US Department of Energy through the DOE EPSCoR Implementation Grant no. DE-SC0007074.

**Figures**

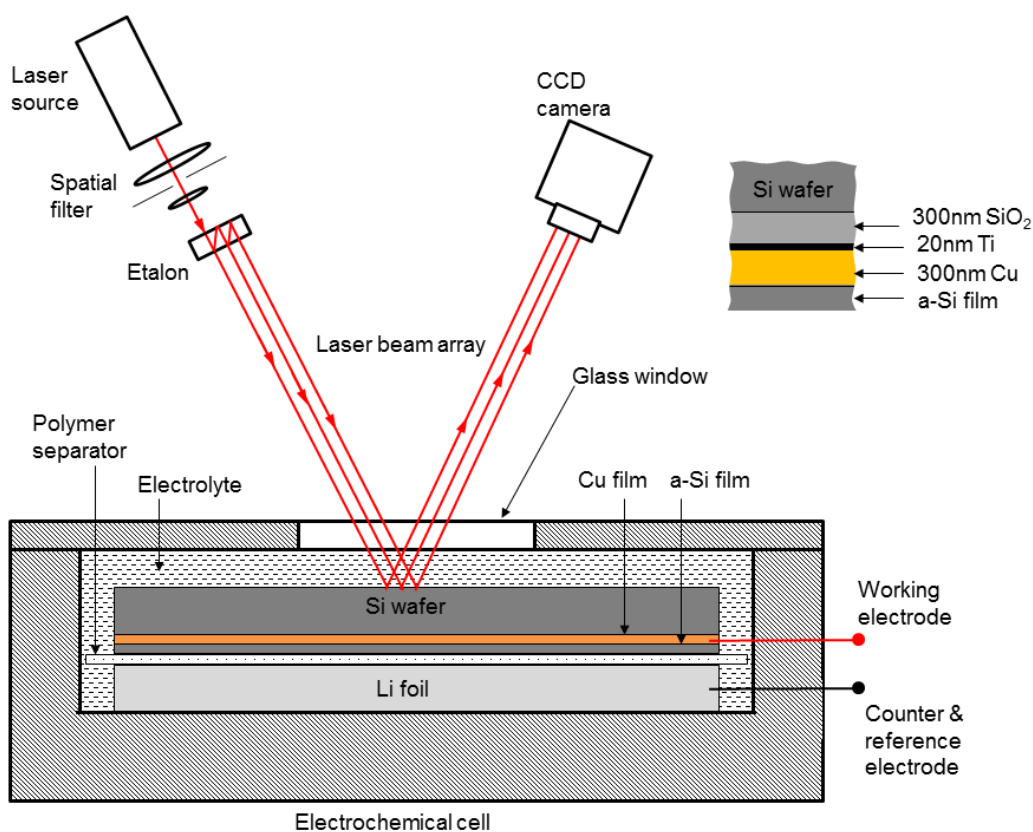

Fig. 1 Schematic illustration of electrochemical cell and the multi-beam optical sensor setup for curvature measurements. The inset shows the details of different films grown and deposited on Si wafer. Figure is not to scale.

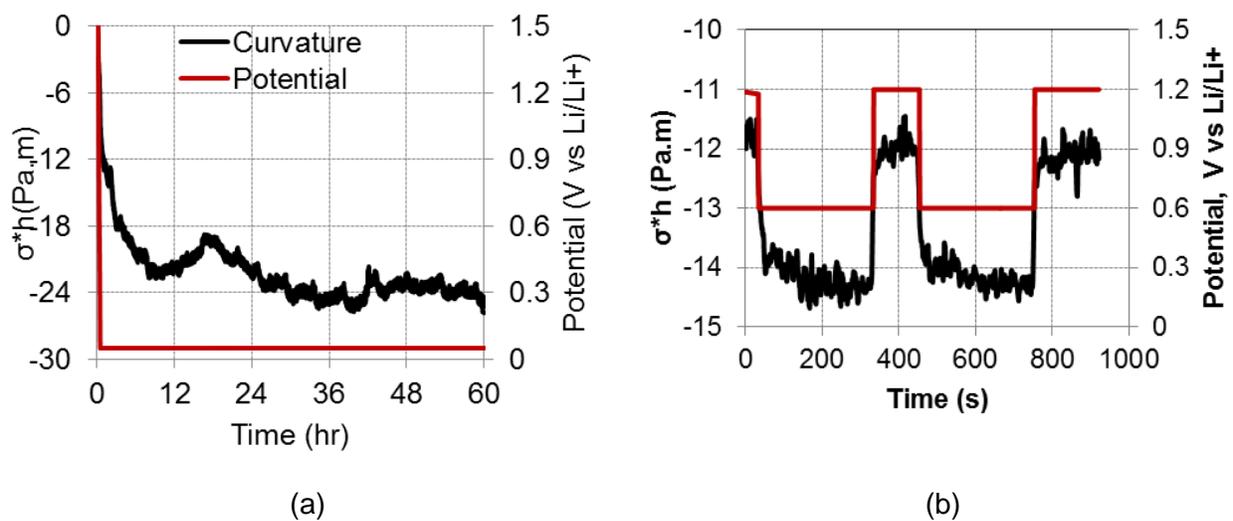

Fig. 2 (a) stress-thickness (which is proportional to substrate curvature ) evolution as a function of time during a 0.05 V vs Li/Li$^+$ potential hold experiment on Cu film sample. (b) The Cu film sample was subjected to instantaneous potential jumps between 0.6 to 1.2 V vs. Li/Li+ to understand the surface charge density induced surface stress contribution to the curvature change.

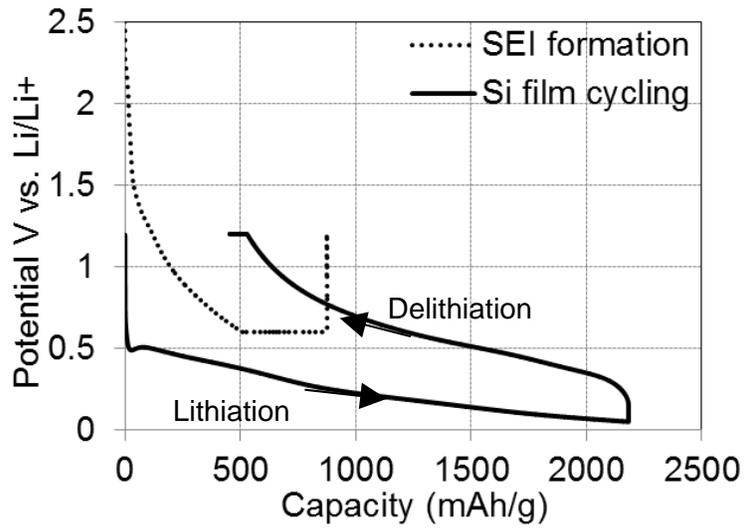

(a)

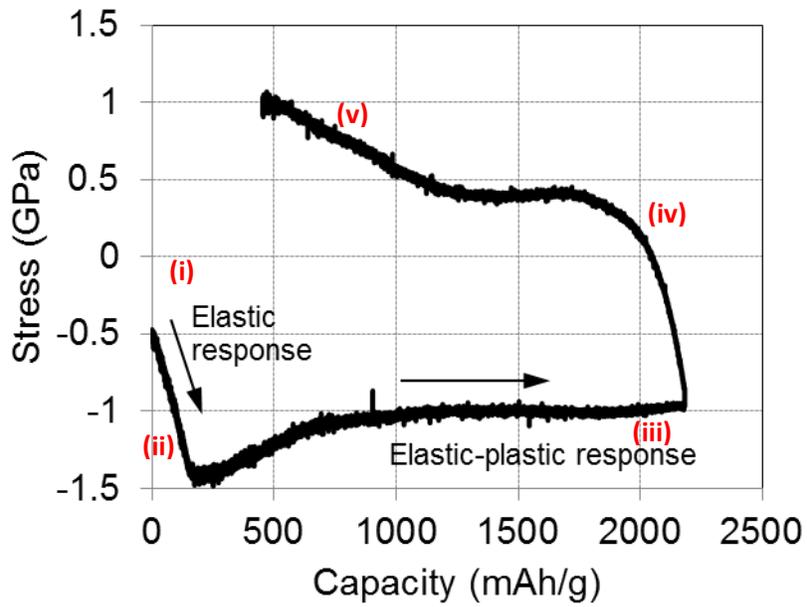

(b)

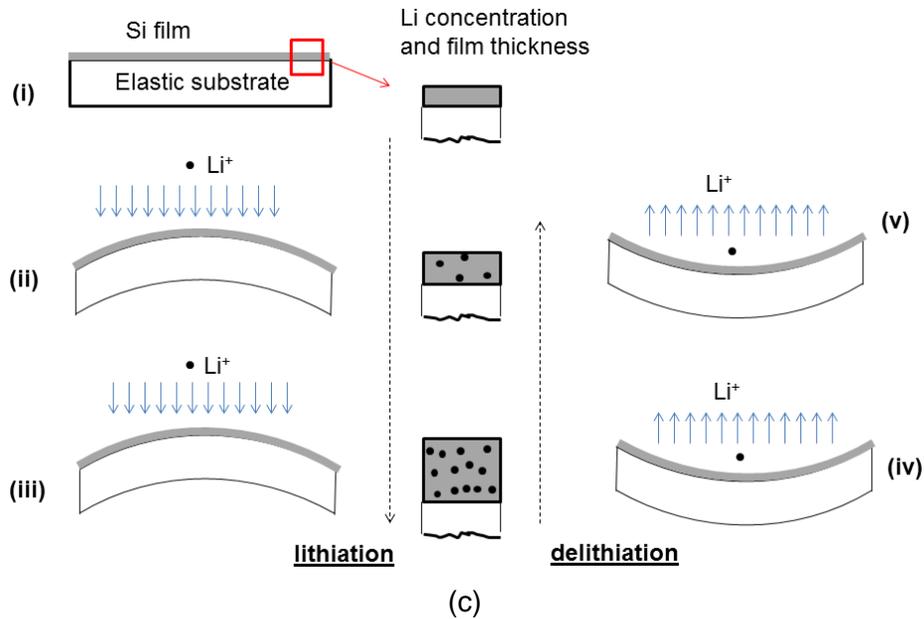

Fig.3 The potential and stress response of 100nm a-Si film (sample 4of Table 1) as a function of capacity (or lithium concentration) is depicted in (a) and (b), respectively. The dashed line in (a) represents the SEI layer formation at 0.6 V *vs.* $Li/Li^+$, and the solid line shows the subsequent lithiation of Si film. The stress measurement during this lithiation is showed in (b). This procedure partially accounts for SEI contribution to the charge during stress measurements. (c) Schematic illustration showing the evolution of curvature, Li concentration, and film thickness at instances (i) – (v) indicated in (b) during lithiation and delithiation; the curvature is exaggerated for clarity. During lithiation, the substrate confinement to in-plane film expansion results in compressive stress and plastic deformation to accommodate volume expansion. During delithiation, the substrate confinement to in-plane contraction results in tensile stress and reverse plastic deformation to accommodate volume contraction.

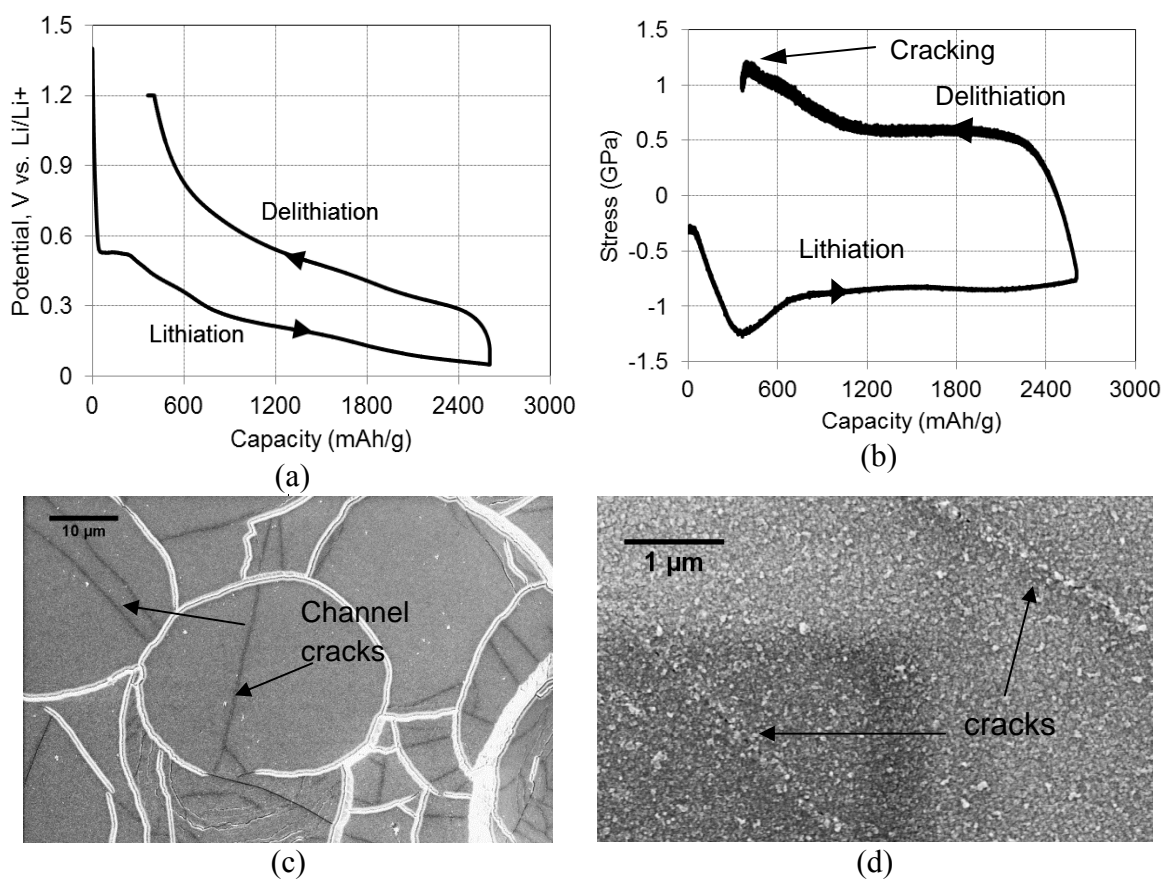

Fig. 4 The potential and stress response of 200nm a-Si film (sample 1of Table 1) as a function of cell capacity (or lithium concentration) in (a) and (b), respectively. SEM image of the film after one cycle of lithiation and delithiation shows (c) channel cracks in the film and (d) surface morphology (roughness) of the film.

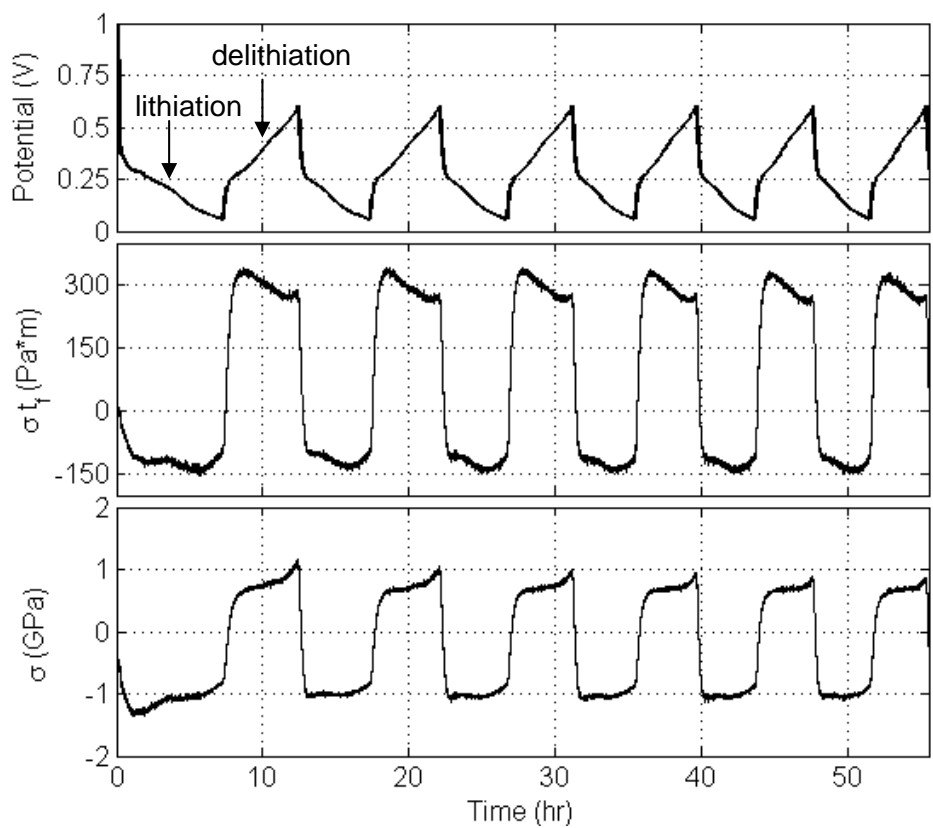

Fig. 5 Experimental data (of sample 3) for the first six cycles of lithiation-delitiation showing (a) Potential, (b) stress multiplied by the film thickness (which is proportional to curvature), and (c) the film stress as a function of time. The film was cycled at C/8 rate under galvanostatic conditions of 8.75 μA/cm$^2$.

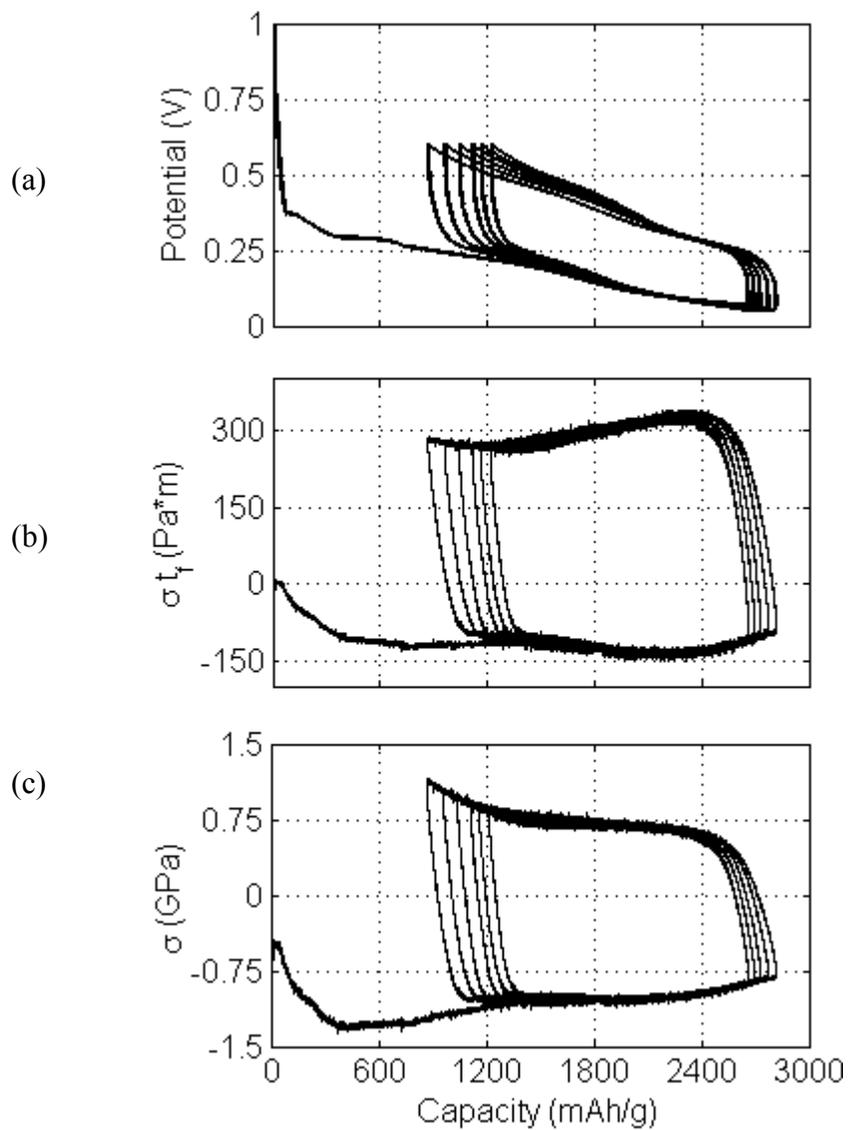

Fig. 6 Electrochemical and mechanical behavior of 100nm Si film during lithiation and delithiation cycling. (a) Potential, (b) Curvature, and (c) film stress as a function of capacity during first six cycles.

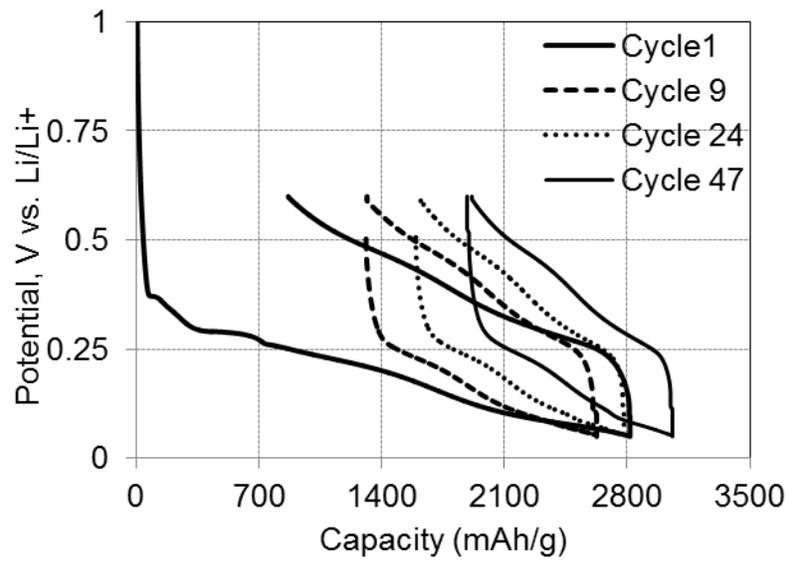

(a)

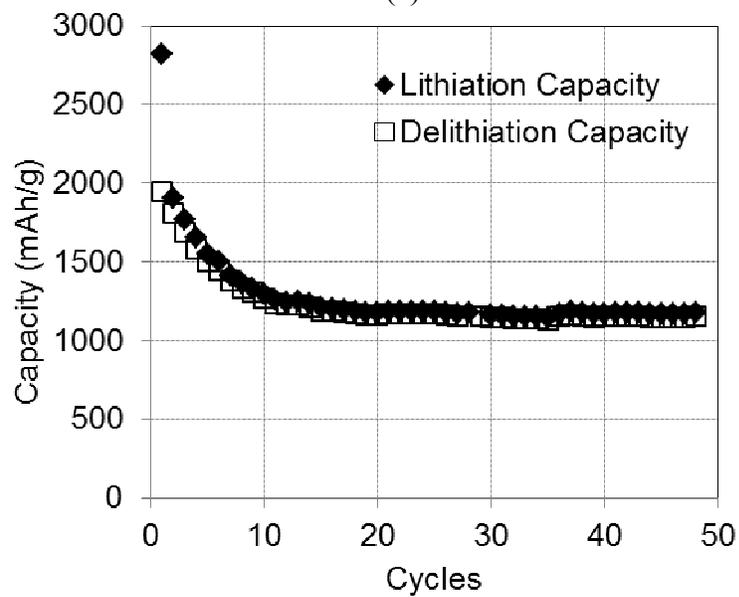

(b)

Fig. 7 Electrochemical data of 100nm a-Si film (Sample 3 of Table 1) cycled under galvanostatic conditions at C/8 rate. (a) Potential as a function of capacity for different cycles. (b) Capacity as a function of cycle number showing lithiation capacity (solid diamond), delithiation capacity (square).

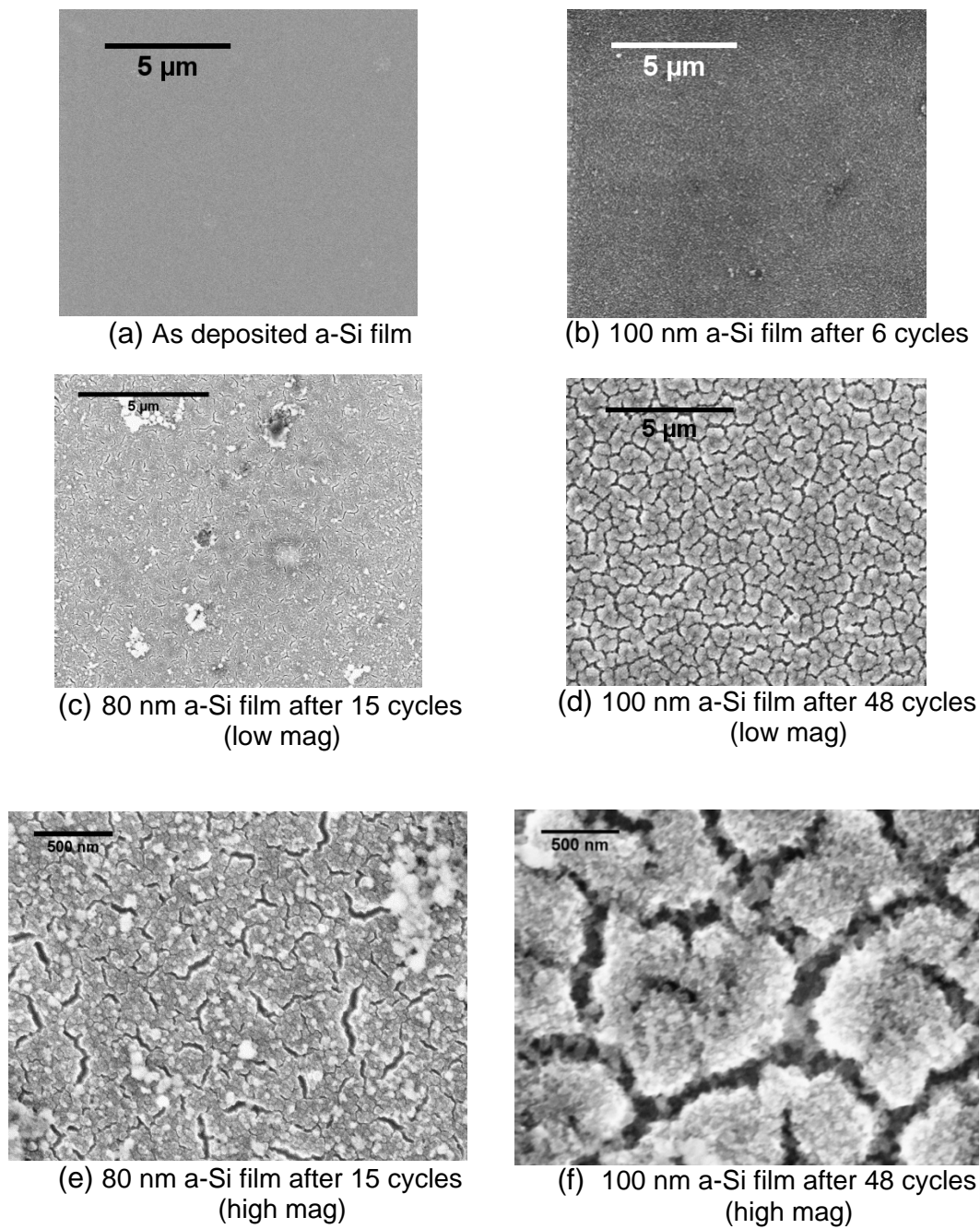

Fig. 8 SEM image of (a) as deposited 100nm thick a-Si film, (b) 100nm film (of Sample 2) after 6 cylces of lithiation/delithation showing no cracks, (c) 80 nm film after 15 cycles showing some cracks, (d)100nm film (of Sample3) after 48 cycles showing extensive cracking and separation into islands. (e) and (f) are high magnification images of (c) and (d), respectively.

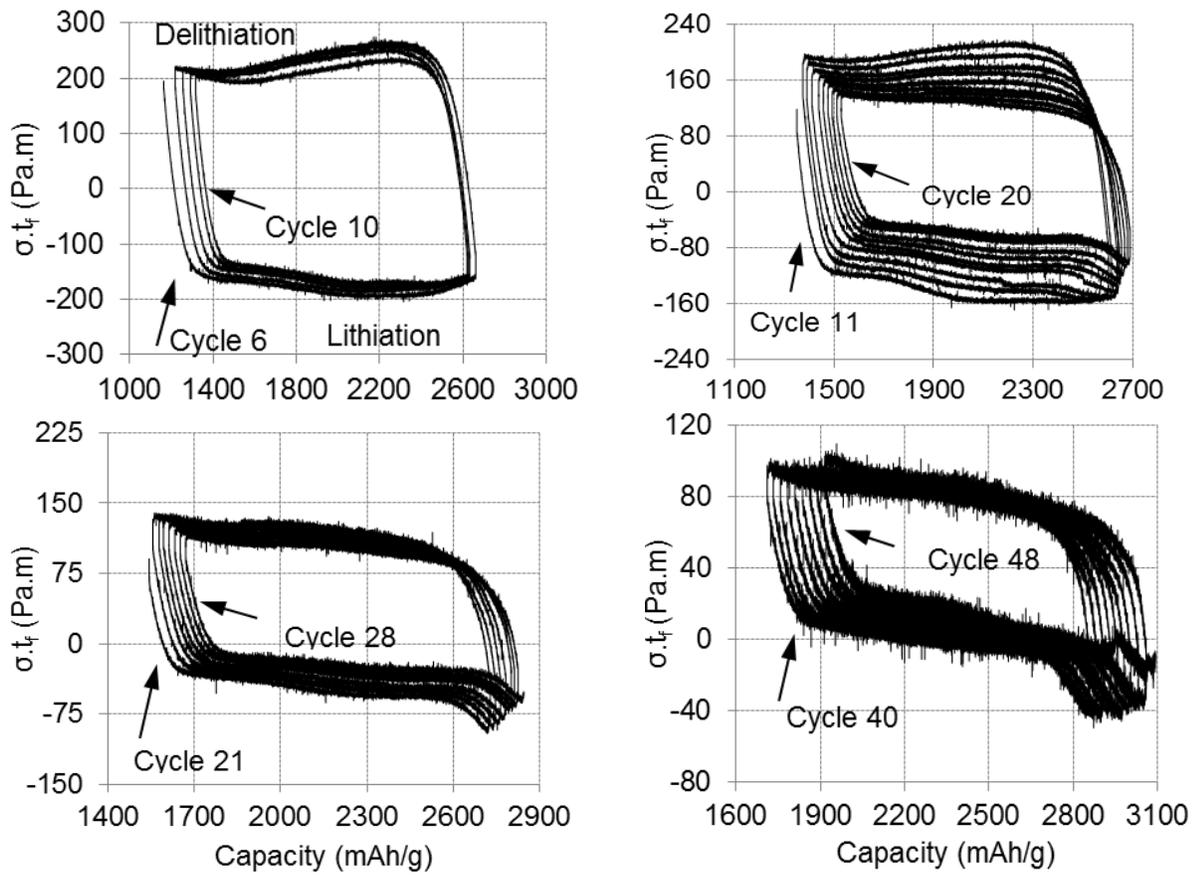

Fig. 9 Evolution of substrate curvature due to electrochemical cycling of 100 nm Si film (of sample 3 of Table 1) as a function of capacity.

# Tables

Table 1 Representative sample geometry and experimental parameters on which the discussion is focused in this manuscript.

| Sample | Electrode configuration | Substrate thickness $t_s$ (μm) | Residual stress in $a$-Si film $\sigma_r$ (GPa) | Cut off voltages (V vs. Li/Li$^+$) | | Number of cycles |
| --- | --- | --- | --- | --- | --- | --- |
| | | | | Lithiation | Delithiation | |
| 1 | 200 nm $a$-Si film on SSP wafer | 298 | -0.29 | 0.05 | 1.2 | 1 |
| 2 | 100 nm $a$-Si film on DSP wafer | 403 | -0.55 | 0.05 | 0.6 | 6 |
| 3 | 100 nm $a$-Si film on DSP wafer | 400 | -0.52 | 0.05 | 0.6 | 48 |
| 4 | 100 nm $a$-Si film on DSP wafer | 487 | -0.46 | SEI layer was grown first followed by a-Si film lithiation and delithiation between 0.05 and 1.2 V vs. Li/Li$^+$ | | |

Table 2 Mechanical properties of different materials used in the study

| Parameter | Definition | Value | Comments |
| --- | --- | --- | --- |
| $E_s$ | Young's modulus of (111) Si wafer substrate | 169 GPa | Ref [35] |
| $E_{si}$ | Young's modulus of $a$-Si film | 80 GPa | Ref [25] |
| $v_s$ | Poisson ratio of (111) Si wafer substrate | 0.26 | Ref [35] |
| $v_{si}$ | Poisson ratio $a$-Si film | 0.22 | Ref [25] |

Table 3 A comparison of fracture resistance of lithiated a-Si with the Mode-I fracture toughness of crystalline Si. The third column shows the data calculated from the relation $\Gamma = \frac{K_{Ic}^2 (1-v^2)}{E}$, where E, υ of (111) Si wafer from Table 2 was used.

| Young's modulus of Li$_{0.4}$Si used in $G$ calculation (GPa) | Estimated fracture energy of Li$_{0.4}$Si (J/m$^2$) | Mode-I fracture energy of crystalline Si with {110} as cleavage plane, $\Gamma$, J/m$^2$ ($K_{Ic}$, MPa√m) |
| --- | --- | --- |
| 52 Ref.[16] | 11 | 13.8 (1.58) Ref.[37], |
| 65 Ref.[36] | 9 | 5 (0.96) Ref.[38], |
| 62 Ref.[20] | 9.4 | 8.9 (1.27) Ref.[39] |

**Figure captions**

1. Schematic illustration of electrochemical cell and the multi-beam optical sensor setup for curvature measurements. The inset shows the details of different films grown and deposited on Si wafer. Figure is not to scale.
2. (a) stress-thickness (which is proportional to substrate curvature ) evolution as a function of time during a 0.05 V vs. Li/Li$^+$ potential hold experiment on Cu film sample. (b) The Cu film sample was subjected to instantaneous potential jumps between 0.6 to 1.2 V vs. Li/Li+ to understand the surface charge density induced surface stress contribution to the curvature change.
3. The potential and stress response of 100nm a-Si film (sample 4of Table 1) as a function of capacity (or lithium concentration) is depicted in (a) and (b), respectively. The dashed line in (a) represents the SEI layer formation at 0.6 V *vs.* Li/Li$^+$, and the solid line shows the subsequent lithiation of Si film. The stress measurement during this lithiation is showed in (b). This procedure partially accounts for SEI contribution to the charge during stress measurements. (c) Schematic illustration showing the evolution of curvature, Li concentration, and film thickness at instances (i) – (v) indicated in (b) during lithiation and delithiation; the curvature is exaggerated for clarity. During lithiation, the substrate confinement to in-plane film expansion results in compressive stress and plastic deformation to accommodate volume expansion. During delithiation, the substrate confinement to in-plane contraction results in tensile stress and reverse plastic deformation to accommodate volume contraction.

4. The potential and stress response of 200nm a-Si film (sample 1of Table 1) as a function of cell capacity (or lithium concentration) in (a) and (b), respectively. SEM image of the film after one cycle of lithiation and delithiation shows (c) channel cracks in the film and (d) surface morphology (roughness) of the film.
5. Experimental data (of sample 3) for the first six cycles of lithiation-delithiation showing (a) Potential, (b) stress multiplied by the film thickness (which is proportional to curvature), and (c) the film stress as a function of time. The film was cycled at C/8 rate under galvanostatic conditions of 8.75 µA/cm$^2$.
6. Electrochemical and mechanical behavior of 100nm Si film during lithiation and delithiation cycling. (a) Potential, (b) Curvature, and (c) film stress as a function of capacity during first six cycles.
7. Electrochemical data of 100nm a-Si film (Sample 3 of Table 1) cycled under galvanostatic conditions at C/8 rate. (a) Potential as a function of capacity for different cycles. (b) Capacity as a function of cycle number showing lithiation capacity (solid diamond), delithiation capacity (square).
8. SEM image of (a) as deposited 100nm thick a-Si film, (b) 100nm film (of Sample 2) after 6 cylces of lithiation/delithiation showing no cracks, (c) 80 nm film after 15 cycles showing some cracks, (d)100nm film (of Sample3) after 48 cycles showing extensive cracking and separation into islands. (e) and (f) are high magnification images of (c) and (d), respectively.
9. Evolution of substrate curvature due to electrochemical cycling of 100 nm Si film (of sample 3 of Table 1) as a function of capacity.